\documentclass[]{article}  

\usepackage[]{graphicx}
\usepackage{amsmath}
\usepackage{amssymb}
\usepackage[]{graphicx}
\usepackage{epsfig}
\usepackage{color}

\def\##1{{\bf #1}}
\def\=#1{\underline{\underline #1}}

\def\.{\mbox{ \tiny{$^\bullet$} }}

\def\les{\left[}
\def\ris{\right]}

\def\c#1{\cite{#1}}

\def\epso{\epsilon_{\scriptscriptstyle 0}}
\def\lambdao{\lambda_{\scriptscriptstyle 0}}
\def\muo{\mu_{\scriptscriptstyle 0}}
\def\ko{k_{\scriptscriptstyle 0}}
\def\etao{\eta_{\scriptscriptstyle 0}}

\def\eps{\epsilon}
\def\epsa{\epsilon_a}
\def\epsb{\epsilon_b}
\def\epsc{\epsilon_c}
\def\epsdt{\epsilon_d}
\def\epsmet{\eps_{met}}

\def\chiv{\chi_v}

\def\ux{\hat{\#u}_x}
\def\uy{\hat{\#u}_y}
\def\uz{\hat{\#u}_z}

\def\tildeQ{{\tilde Q}}

\begin{document}
\noindent{\bf Multiple trains of same-color surface  plasmon-polaritons guided by the planar
interface of a metal and a sculptured nematic thin film.\\ Part IV: Canonical problem}\\


\noindent{Muhammad Faryad,$^a$ John A. Polo Jr.,$^b$ and Akhlesh Lakhtakia$^{a,c}$}\\

\noindent$^a$Nanoengineered Metamaterials Group (NanoMM),
Department of Engineering Science and
Mechanics,
Pennsylvania State University, University Park, PA  16802-6812, USA\\

\noindent$^{b}$Department of Physics and Technology, Edinboro University of Pennsylvania, 235 Scotland Rd., Edinboro, PA  16444, USA\\

\noindent$^c$Department of Physics, Indian Institute of Technology Kanpur, Kanpur (UP) 208016, India\\

\noindent {\bf Abstract.}
The canonical problem of the propagation of surface-plasmon-polariton (SPP) waves localized to the
planar interface of a metal and a  sculptured nematic thin film (SNTF) that
is periodically nonhomogeneous along the direction normal to the interface was formulated.
Solution of the dispersion equation obtained thereby confirmed the possibility of exciting
multiple SPP waves of the same frequency or color. However, these SPP waves differ
in phase speed, field structure, and the e-folding distance along the direction
of propagation.

\newpage
\section{Introduction}

Among the various forms of electromagnetic surface waves, the surface-plasmon-polariton~(SPP) wave has the longest history of theoretical development and application.  Arising from earlier developments, 
 the SPP wave was envisioned in the mid-twentieth century as a wave guided by the planar interface of a metal and an isotropic dielectric material. Since that time, the theory has evolved to encompass interfaces between a metal and various dielectric materials of greater complexity.  The inclusion of anisotropic, homogeneous dielectrics\cite{Borstel,Wallis,Elston,Depine95,Wang,Depine97,Yan,Abdulhalim} in the study of electromagnetic surface waves has been considered for some time now.  More recently, SPP waves at the interface of a metal and a nonhomogeneous dielectric material, which exhibit several interesting properties, has been receiving considerable attention. The nonhomogeneous dielectric materials investigated include continuously varying materials\cite{Sprokel_I,Sprokel_II,PL2009,PLjosaa} such as liquid crystals,  as well as layered structures\cite{Gaspar,Das,Guo_I,Guo_II}.

This series of papers is devoted
to theoretical and experimental investigations on the propagation of a surface electromagnetic wave guided
by the planar interface of a metal and a sculptured nematic thin film (SNTF), the latter being periodically nonhomogeneous in the direction normal to the interface.
In Parts I~\cite{PartI} and II~\cite{PartII}, the absorbance, reflectance and transmittance of an incident  linearly polarized plane wave were calculated when this planar interface is implemented {\em via}
a metal-SNTF bilayer and a prism in a Kretschmann setup. The wavevector of the incident plane wave was supposed to lie
wholly in the morphologically significant plane of the SNTF, but that restriction was lifted in Part~II.  The
computed results showed that  multiple surface-plasmon-polariton~(SPP) waves, all of the same frequency or color,
can be excited
at the metal/SNTF interface. The guided SPP waves possess different field structures   as well as
different phase speeds.

In Part III~\cite{PartIII}, experimentally obtained data was presented in support of the theoretical predictions of Part~I. The absorbances of two different metal-SNTF bilayers  were   evaluated as the difference between unity and the measured reflectance,
the transmittance being assumed to be null-valued at angles of incidence exceeding the critical angle  for
the interface of the SNTF with the prism. Analysis of the collected data
 confirmed the possibility of exciting multiple SPP waves with different phase speeds and field structures.
 Parenthetically, in a parallel effort
 the same possibility was theoretically predicted \cite{PL2009,PLjosaa} and validated experimentally \cite{DPL2009}
 for the planar interface of a metal and a chiral sculptured thin film \cite{Lmsec,LMbook}.

The method adopted for theoretical predictions and experimental verification in Parts I--III was an indirect one, because the existence of SPP waves was deduced from the absorbance, reflectance and transmittance characteristics of a metal-SNTF
bilayer. Each layer in the bilayer is of finite thickness. In the canonical problem of SPP-wave propagation, the metal and the SNTF have to be semi-infinite. The solution of the canonical problem provides incontrovertible proof  of the existence of multiple SPP-wave modes, and also eliminates possible confusion with waveguide modes spread over the entirety of the SNTF.
Accordingly, for this paper,
we set out to prove the existence of multiple SPP waves directly. The approach adopted by us is similar to
that of Agarwal {\it et al.}~\cite{APL2009}, being independent of any incident plane wave. The formulated boundary-value problem   was solved numerically
for the SPP wavenumbers.

The canonical problem is formulated
in Sec.~\ref{theory}, and numerical results are presented
and discussed in Sec.~\ref{nrd}.
An $\exp(-i\omega t)$ time-dependence is implicit, with $\omega$
denoting the angular frequency. The free-space wavenumber, the
free-space wavelength, and the intrinsic impedance of free space are denoted by $\ko=\omega\sqrt{\epso\muo}$,
$\lambdao=2\pi/\ko$, and
$\etao=\sqrt{\muo/\epso}$, respectively, with $\muo$ and $\epso$ being  the permeability and permittivity of
free space. Vectors are in boldface, dyadics are underlined twice,
column vectors are in boldface and enclosed within square brackets, and
matrixes are underlined twice and square-bracketed. The asterisk denotes the complex conjugate,
and the Cartesian unit vectors are
identified as $\ux$, $\uy$, and $\uz$.

\section{Theory}\label{theory}

Let the half-space $z\leq0$ be occupied by an isotropic and homogeneous metal
with
complex-valued relative permittivity scalar $\epsmet$. The region $z\geq0$ is occupied by the chosen SNTF with a
periodically nonhomogeneous  permittivity dyadic \cite{PartI,PartII,PartIII}
\begin{equation}
\=\epsilon_{SNTF}(z)= \epso\, \=S_{y}(z) \cdot \=\epsilon^{\circ} _{ref}(z) \cdot \=S_{y}^{-1}(z)\,,
\end{equation}
where the dyadics
\begin{equation}
\left.\begin{array}{l}
\=S_{y}(z)=(\ux\ux+\uz\uz) \cos\left[\chi (z)\right] +
(\uz\ux - \ux\uz) \sin \left[\chi (z)\right] +\uy\uy
\\[5pt]
\=\epsilon^{\circ} _{ref}(z)=\epsa(z)\,\uz\uz+\epsb(z)\,\ux\ux+\epsc(z)\,\uy\uy
\end{array}\right\}
\end{equation}
depend on the vapor incidence angle
$\chiv(z) = {\tilde\chi}_v + \delta_v\,\sin(\pi z/\Omega)$
that varies sinusoidally with period $2\Omega$.

In order to investigate SPP-wave propagation, we adopted a procedure devised
to investigate the propagation of Dyakonov--Tamm waves \cite{APL2009}.
Let the SPP wave propagate parallel to the unit vector $\ux\cos\psi+\uy\sin\psi$
along the interface $z=0$, and attenuate as
$z\to\pm\infty$.
Therefore, in the region $z \leq 0$, the wave vector may be written as
\begin{equation}
\#k_{met}=\kappa\, \hat{\#u}_1 -\alpha_{met}\,\uz\,,
\end{equation}
where
$\kappa^2+\alpha_{met}^2=\ko^2\, \epsmet$,
 $\kappa$ is complex-valued, and
 ${\rm Im}(\alpha_{met})>0$ for attenuation as $z\to-\infty$; here and hereafter,
the unit vectors
$\hat{\#u}_1= \ux \cos\psi + \uy \sin\psi$ and
$\hat{\#u}_2= - \ux \sin\psi + \uy \cos\psi$.
Accordingly, the field phasors in the metal may be written as
\begin{equation}
\#E(\#r)= \left[a_{p}\left( \frac{\alpha_{met}}{\ko}\,\hat{\#u}_1+\frac{\kappa}{\ko}\, \uz\right)+a_{s}\,\hat{\#u}_2\right]
\exp(i\#k_{met}\cdot\#r)\,,\quad z \leq 0\,,
\label{eqn:Esub}
\end{equation}
and
\begin{equation}
\#H(\#r)=\etao^{-1}\left[  -
a_{p}\,\epsilon_{met}\,\hat{\#u}_2+a_{s} \left(\frac{\alpha_{met}}{\ko}\,\hat{\#u}_1+\frac{\kappa}{\ko}\,\uz\right)\right] \exp(i\#k_{met}\cdot\#r)\,,\quad z \leq 0\,,\label{eqn:Hsub}
\end{equation}
where $a_{p}$ and $a_{s}$ are unknown scalars.

 For field representation in the SNTF, let us
write
$\#E(\#r)=\#e(z)\,\exp\left( {i\kappa\hat{\#u}_1\cdot\#r}\right)$ and
$\#H(\#r)=\#h(z)\,\exp\left( {i\kappa\hat{\#u}_1\cdot\#r}\right)$.
The components $e_z(z)$ and $h_z(z)$ of the field phasors can be found in terms of
the other components as follows:
\begin{eqnarray}
e_z(z)&=&\frac{\epsdt(z)\left[\epsa(z)-\epsb(z)\right]\sin\left[\chi(z)\right]\cos\left[\chi(z)\right]}{\epsa(z)\epsb(z)}e_x(z)\nonumber\\&&+\kappa\frac{\epsdt(z)}{\omega\epso\epsa(z)\epsb(z)}\left[h_x(z)\sin\psi - h_y(z)\cos\psi\right],\quad z > 0\,, \\
h_z(z)&=&-\,\frac{\kappa}{\omega\muo}\left[ e_x(z)\sin\psi- e_y(z)\cos\psi\right],\quad z > 0\,,
\end{eqnarray}
where
\begin{equation}
\epsdt(z)=  \frac{\epsa(z)\,\epsb(z)}{
\epsa(z)\,\cos^2\les\chi(z)\ris + \epsb(z)\,\sin^2\les\chi(z)\ris}
\,.
\end{equation}

The other components of the electric and magnetic field phasors are used in the column vector
\begin{equation}
\left[\#f(z)\right]= \left[e_x(z)\quad e_y(z) \quad h_x(z)\quad h_y(z)\right]^T\,
\end{equation}
which satisfies the matrix differential equation \cite{APL2009}
\begin{equation}
\label{MODE}
\frac{d}{dz}\left[\#f(z)\right]=i \left[\=P(z)\right]\cdot\left[\#f(z)\right]\,,
\quad z>0\,,
\end{equation}
where the 4$\times$4 matrix
\begin{eqnarray}
\nonumber
&&
[\=P(z)]=
\omega\,\les\begin{array}{cccc}
0 & 0 & 0 & \muo \\
0 & 0 & -\muo & 0 \\
0 & -\epso\,\epsc(z) & 0 & 0\\
\epso\,\epsdt(z) & 0 & 0 & 0
\end{array}\ris
\\[4pt]
\nonumber
&& \qquad
+\,\kappa\,\frac{\epsdt(z)\,\les\epsa(z)-\epsb(z)\ris}{\epsa(z)\,\epsb(z)}\,
\sin\les\chi(z)\ris\,\cos\les\chi(z)\ris\,
\les\begin{array}{cccc}
\cos\psi & 0 & 0 & 0\\
\sin\psi & 0 & 0 & 0\\
0 & 0 & 0 & 0\\
0 & 0 & -\sin\psi & \cos\psi
\end{array}\ris
\\[4pt]
\nonumber
&& \qquad\quad
+\,\frac{\kappa^2}{\omega\epso}\,
\frac{\epsdt(z)}{\epsa(z)\,\epsb(z)}\,
\les\begin{array}{cccc}
0 & 0 & \cos\psi\,\sin\psi & -\cos^2\psi \\
0 & 0 & \sin^2\psi & -\cos\psi\,\sin\psi \\
0 & 0 & 0 & 0\\
0 & 0 & 0 & 0
\end{array}\ris
\\[4pt]
\label{eq7.14}
&&\qquad\qquad
+\,\frac{\kappa^2}{\omega\muo}\,
\les\begin{array}{cccc}
0 & 0 & 0 & 0\\
0 & 0 & 0 & 0\\
-\cos\psi\,\sin\psi & \cos^2\psi & 0 & 0\\
-\sin^2\psi & \cos\psi\,\sin\psi & 0 & 0
\end{array}\ris\,.
\end{eqnarray}

As in Part~II, we used the piecewise uniform approximation technique \cite{LMbook}
to determine the matrix $[\=Q]$ that appears in the relation
\begin{equation}
[\#f(2\Omega)]=[\=Q]\cdot[\#f(0+)]
\end{equation}
to characterize the optical response of one period of the chosen SNTF for specific
values of $\kappa$ and $\psi$. By virtue of the Floquet--Lyapunov theorem
\c{YS75},  a matrix $[\=\tildeQ]$ can be defined such that
\begin{equation}
[\=Q] = \exp\left\{i2\Omega[\=\tildeQ]\right\}\,.
\end{equation}
Both $[\=Q]$ and $[\=\tildeQ]$ share the same eigenvectors, and their eigenvalues
are also related. Let $[\#t]^{(n)}$, $n\in\left[1,4\right]$, be the eigenvector corresponding
to the  $n$th eigenvalue $\sigma_n$ of $[\=Q]$; then, the corresponding eigenvalue
$\alpha_n$ of $[\=\tildeQ]$
is given by
\begin{equation}
\alpha_n = -i\frac{\ln \sigma_n}{2\Omega}\,,\qquad n\in\left[1,4\right]\,.
\end{equation}

After ensuring that ${\rm Im}({\alpha_{1,2}})>0$, we set
\begin{equation}
[\#f(0+)]= \left[\,[\#t]^{(1)}\quad [\#t]^{(2)}\,\right]\cdot
\left[\begin{array}{c} b_1\\ b_2\end{array}\right]\,
\end{equation}
for SPP-wave propagation,
where $b_1$ and $b_2$ are unknown scalars;
the other two eigenvalues of $[\=\tildeQ]$ pertain to waves that amplify as $z\to\infty$
and cannot therefore contribute to the SPP wave.
At the same time,
\begin{equation}
[\#f(0-)]=\left[\begin{array}{cc}
\frac{\alpha_{met}}{\ko}\,\cos\psi &-\sin\psi \\[5pt]
\frac{\alpha_{met}}{\ko}\,\sin\psi&\cos\psi\\[5pt]
\frac{\epsilon_{met}}{\etao}\,\sin\psi&\frac{\alpha_{met}}{\ko\etao}\,\cos\psi\\[5pt]
-\frac{\epsilon_{met}}{\etao}\,\cos\psi&\frac{\alpha_{met}}{\ko\etao}\,\sin\psi
\end{array}\right]\cdot
\left[\begin{array}{c} a_p\\ a_s\end{array}\right]\,,
\end{equation}
by virtue of (\ref{eqn:Esub}) and (\ref{eqn:Hsub}). Continuity of the tangential
components of the electric and magnetic field phasors across the plane
$z=0$ requires that
$[\#f(0-)]=[\#f(0+)]$,
which may be rearranged as the matrix equation
\begin{equation}
[\=Y]\cdot\left[\begin{array}{c}a_p\\ a_s\\ b_1\\ b_2\end{array}\right]=
\left[\begin{array}{c}0\\0\\0\\0\end{array}\right]\,.
\label{eq:tobsol}
\end{equation}
For a nontrivial solution, the 4$\times$4 matrix $[\=Y]$ must be singular,
so that
\begin{equation}
{\rm det}\,[\=Y]= 0
\label{eq:SPPdisp}
\end{equation}
is the dispersion equation for the SPP wave. This equation has to be solved in
order to determine the SPP wavenumber $\kappa$.

\section{Numerical Results and Discussion}\label{nrd}

A Mathematica\texttrademark~program
was written and implemented to solve (\ref{eq:SPPdisp}) using the
Newton--Raphson method
\cite{Jaluria} to obtain $\kappa$ for a specific value of $\psi$.
The free-space wavelength was fixed at $\lambdao=633$~nm. The metal was taken to be  aluminum: $\epsmet=-56+21i$.
The SNTF was chosen
to be made of titanium oxide \cite{HWH}, with
\begin{equation}
\left.\begin{array}{l}
\epsa(z)=[1.0443+2.7394 v(z)-1.3697 v^2(z)]^2\\[5pt]
\epsb(z)=[1.6765+1.5649 v(z)-0.7825 v^2(z)]^2\\[5pt]
\epsc(z)=[1.3586+2.1109 v(z)-1.0554 v^2(z)]^2\\[5pt]
\chi(z)=\tan^{-1}[2.8818\tan\chiv(z)]
\end{array}\right\}
\end{equation}
where  $v(z)=2\chiv(z)/\pi$.
The angles $\tilde\chi_v$ and $\delta_v$ were taken to be 45$^\circ$ and 30$^\circ$, respectively,
and $\Omega=200$~nm
for all results presented here.   

\begin{figure}[!ht]
\begin{center}$
\begin{array}{cc}
\includegraphics[width=1.75in]{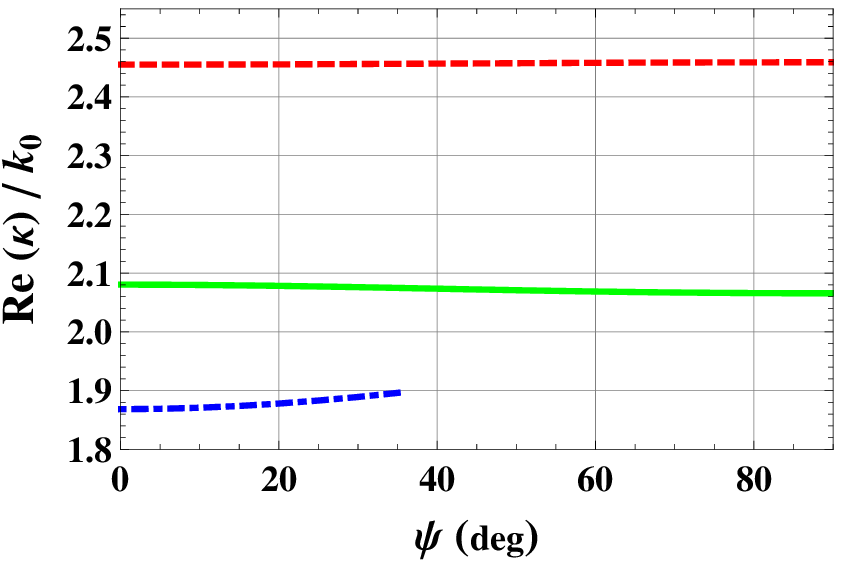} &
\includegraphics[width=1.75in]{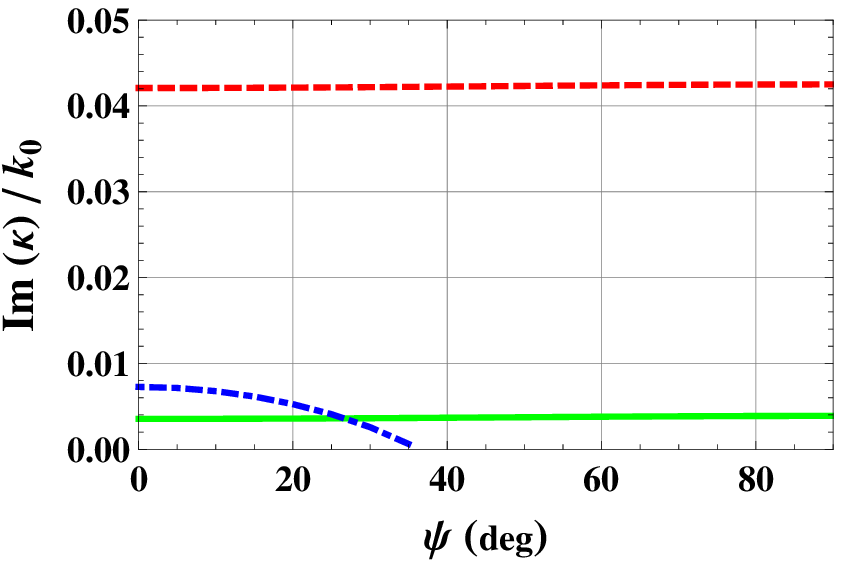}
\end{array}$
\caption{(left) Real and (right) imaginary parts of $\kappa$ as functions of $\psi$, for SPP-wave propagation
guided by the planar interface of aluminum and a titanium-oxide SNTF. See the text for the
constitutive parameters used.  Either two or three modes are possible, depending on $\psi$.}
\label{kappa}
\end{center}
\end{figure}

 The wavenumber $\kappa$ must be complex-valued for the canonical problem
 but only real-valued for the Kretschmann configuration \cite{PL2009}.
Computed values of the
 real and imaginary parts of $\kappa$ for the canonical
 problem are shown in Fig.~\ref{kappa}. For $0^\circ\leq \psi \lesssim 36^\circ$, we found three values of $\kappa$ which satisfy  (\ref{eq:SPPdisp}) and therefore represent SPP waves. For $36^\circ\lesssim \psi \leq 90^\circ$, there are two values of $\kappa$ which satisfy  (\ref{eq:SPPdisp}). This trend is fully
consistent with the conclusions drawn in Part~II for the Kretschmann configuration. The different solutions of
(\ref{eq:SPPdisp}) for any specific value of $\psi$ indicate that the SPP waves have different phase speeds
$ \omega/{\rm Re}(\kappa)$---as theoretically predicted in Parts~I and II, and experimentally
confirmed in Part~III---and different e-folding distances $1/{\rm Im}(\kappa)$ along the direction of propagation.

\begin{figure}[!ht]
\begin{center}
\includegraphics[width=3.6in]{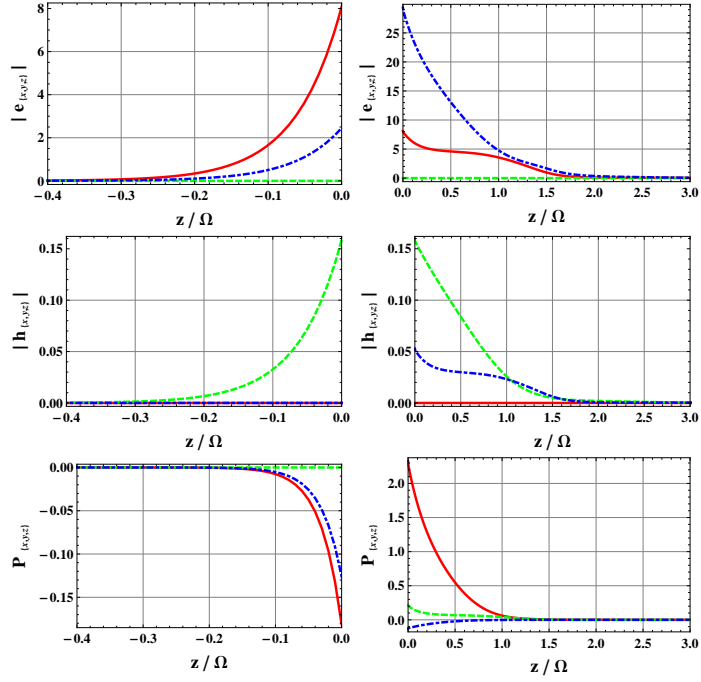}
\caption{Variations of components of $\#e$, $\#h$, and $\#P$ with $z$, for $\kappa=(2.455 +i 0.04208)\ko$ and $\psi = 0^\circ$.
The $x$-, $y$-, and $z$-directed components are represented by solid, dashed and chain-dashed lines, respectively.
The data were computed by setting $a_p=1$~V~m$^{-1}$, with $a_p$, $b_1$, and $b_2$ then obtained using (\ref{eq:tobsol}).}
\label{plot0_1}
\end{center}
\end{figure}

\begin{figure}[!ht]
\begin{center}
\includegraphics[width=3.6in]{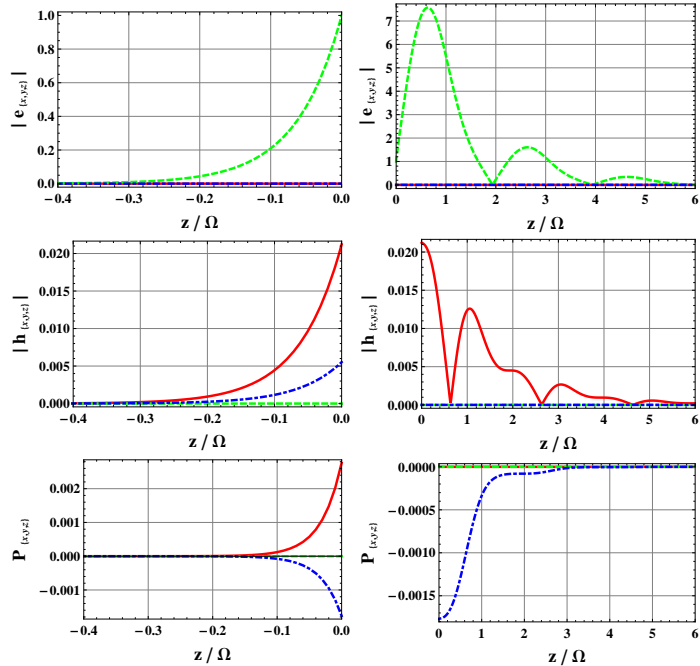}
\caption{Same as Fig.~\ref{plot0_1} except for $\kappa=(2.080 + i 0.003538)\ko$.
The data were computed by setting $a_s=1$~V~m$^{-1}$, with $a_s$, $b_1$, and $b_2$ then obtained using (\ref{eq:tobsol}).}
\label{plot0_2}
\end{center}
\end{figure}

\begin{figure}[!ht]
\begin{center}
\includegraphics[width=3.6in]{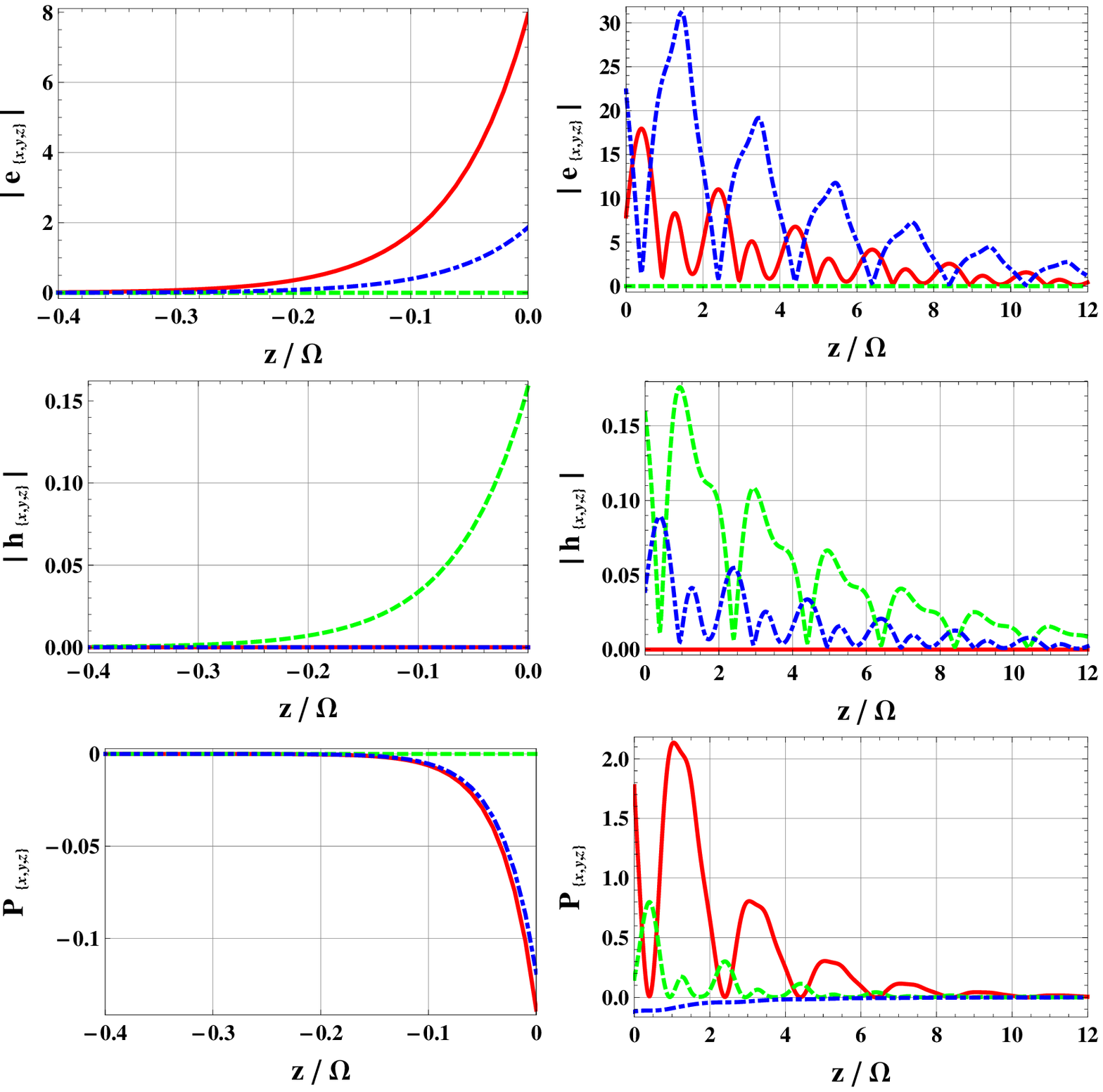}
\caption{Same as Fig.~\ref{plot0_1} except for $\kappa=(1.868 + i 0.007267)\ko$.}
\label{plot0_3}
\end{center}
\end{figure}

 Specifically, for $\psi=0^\circ$ the values of $\kappa$ which satisfy  (\ref{eq:SPPdisp})  are $\kappa_1=(2.455 +i 0.04208)\ko$, $\kappa_2=(2.080 + i 0.003538)\ko$, and $\kappa_3=(1.868 + i 0.007267)\ko$. These solutions represent SPP waves with
 wave vectors lying wholly in the morphologically significant plane of the SNTF, as   addressed theoretically in Part~I and experimentally in Part~III.

 The Cartesian components of the electric and magnetic field phasors and the
  time-averaged Poynting vector $\bf{P}={1\over 2}{\rm Re}\left(\bf{E}\times\bf{H}^\ast\right)$
  as functions of $z$  along the line ($x=0$, $y=0$) are shown for $\kappa_1$  in Fig.~\ref{plot0_1}, and for $\kappa_3$
  in Fig.~\ref{plot0_3}.  These SPP waves are $p$-polarized, and were respectively labeled as $p1$ and $p2$ in Part~III.
  Figure~\ref{plot0_2} shows the variations of $\#e$, $\#h$, and $\#P$ along the $z$ axis for $\kappa=\kappa_2$. This
  SPP wave is $s$-polarized and was labeled as $s3$ in Part~III. The localization of all three SPP waves around the interface
  $z=0$ is evident from Figs.~\ref{plot0_1}--\ref{plot0_3}. Also, the SPP wave $p1$  is more localized
  inside the SNTF than either $p2$ or $s3$. Furthermore, the phase speed
  of $p1$ is higher than that of $s3$, which exceeds the
  phase speed of $p2$. However,  $s3$ will travel a longer distance along the interface than
  either $p1$ or $p2$, which could not have been deduced from the theoretical analysis for
  the Kretschmann configuration in Part~I.

\begin{figure}[!ht]
\begin{center}
\includegraphics[width=3.6in]{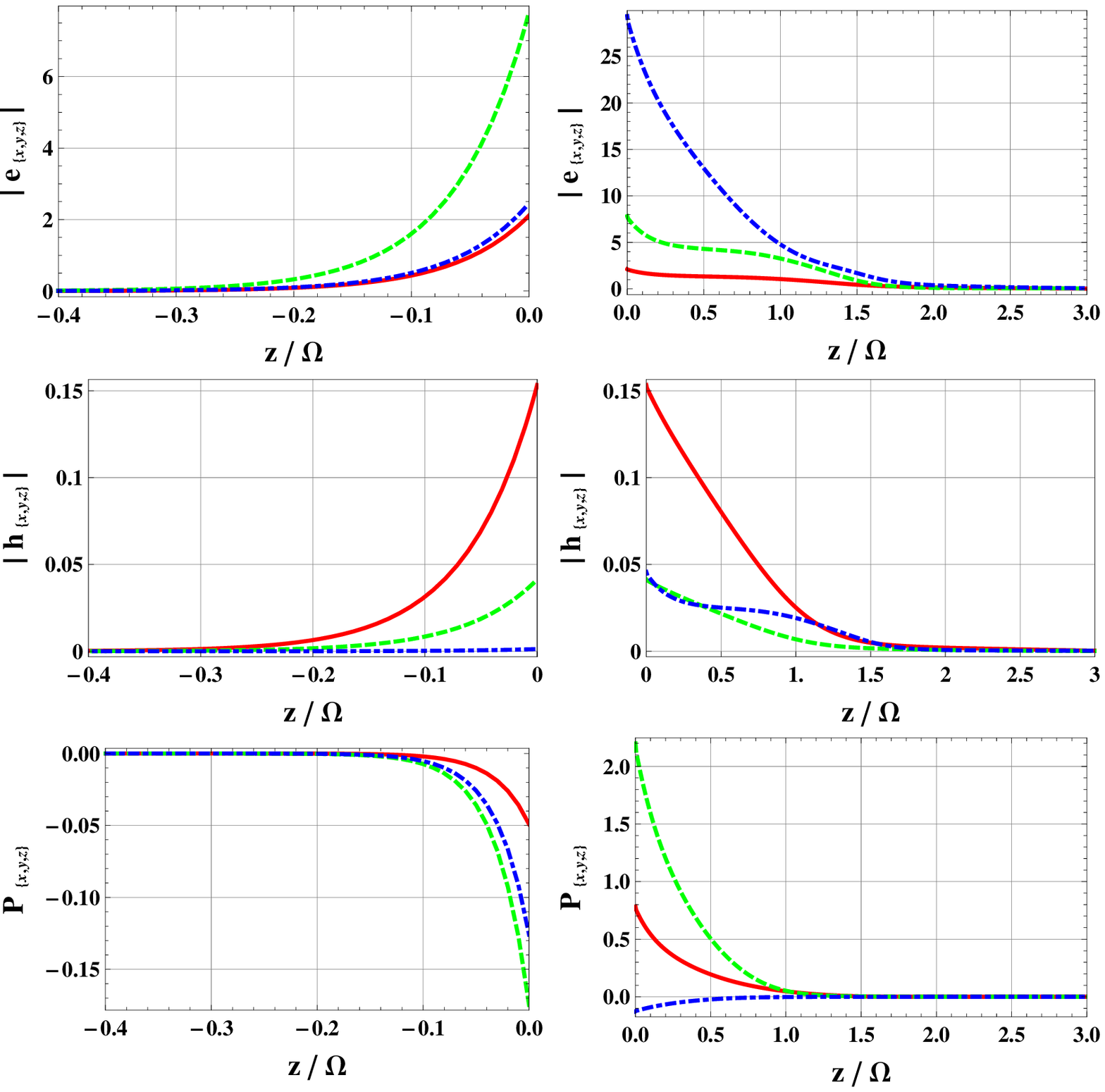}
\caption{Same as Fig.~\ref{plot0_1} except for $\kappa=(2.459 + i 0.04247)\ko$ and $\psi =75^ \circ$.}
\label{plot75_1}
\end{center}
\end{figure}

\begin{figure}[!ht]
\begin{center}
\includegraphics[width=3.6in]{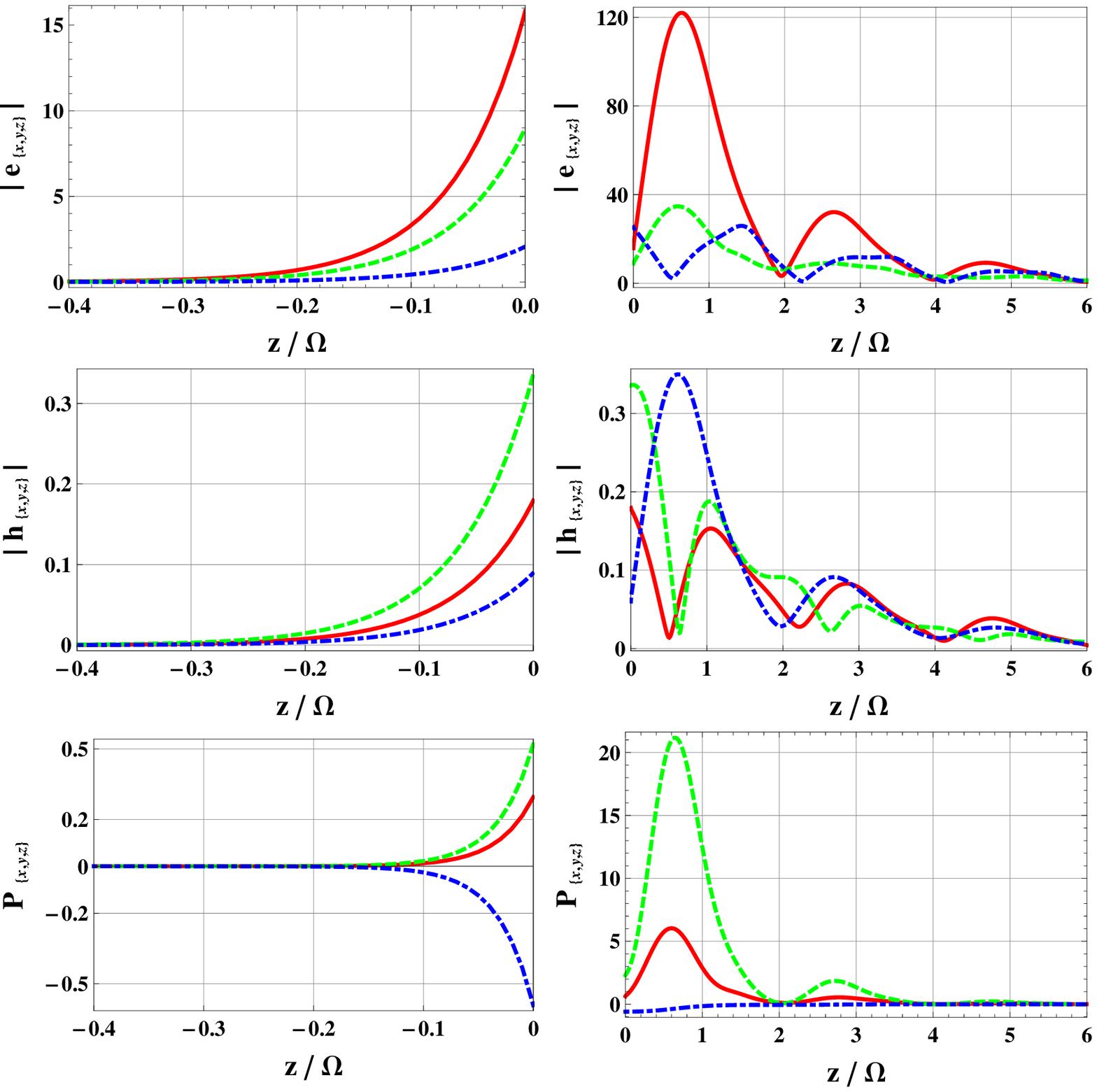}
\caption{Same as Fig.~\ref{plot0_1} except for $\kappa=(2.066 + i 0.003861)\ko$ and $\psi =75^ \circ$.}
\label{plot75_2}
\end{center}
\end{figure}

For $\psi = 75^\circ$, only two values of $\kappa$ were found
to satisfy (\ref{eq:SPPdisp}): $\kappa_1=(2.459 + i 0.04247)\ko$ and $\kappa_2=(2.066 + i 0.003861)\ko$. The variations
of $\#e$, $\#h$, and $\#P$ along the $z$ axis  are shown in Fig.~{\ref{plot75_1} for $\kappa_1$, and  in Fig.~\ref{plot75_2}
for $\kappa_2$. These two SPP waves cannot be classified
as either $p$- or $s$-polarized, which is consistent with the deductions
 in Part~II for the Kretschmann configuration.   Although the
 phase speed of the SPP wave with $\kappa=\kappa_1$ exceeds the phase speed of
 the other SPP wave ($\kappa=\kappa_2$), the latter SPP wave
 will propagate a longer distance along the interface than the former SPP wave.

\section{Concluding Remarks}\label{conc}

In Parts~I and~II of this series, the possibility of excitation of multiple SPP waves guided
by the planar interface of a metal and a periodically
nonhomogeneous SNTF was predicted after solving a boundary-value problem that models
the Kretschmann setup.  In Part~III, an experimental validation was provided for the results of Part~I. However, these approaches  were indirect and a direct proof of the existence of multiple SPP waves was lacking.

The direct proof was provided here after formulating and solving the canonical problem
of propagation guided by the planar interface of semi-infinite expanses of the metal and the
periodically nonhomogeneous SNTF. Not only were the conclusions obtained  in Parts~I--III upheld,
but additional information on the e-folding distance along the direction of propagation also
emerged from the solution of the canonical problem.\\

\noindent {\bf Acknowledgments.}
MF thanks the Trustees of the Pennsylvania State University for a University Graduate Fellowship. AL
is grateful to  Charles Godfrey Binder Endowment
at the Pennsylvania State University for partial support of this work.

 \vspace{2ex}\noindent{\bf Muhammad Faryad} received a B.Sc. degree in Mathematics and Physics from University of Punjab, Lahore, Pakistan, in 2002, and M.Sc. and M.Phil. degrees in Electronics from Quaid-i-Azam University, Islamabad, Pakistan in 2006 and 2008, respectively. His research experience includes analysis of high-frequency fields reflected from cylindrical reflectors in an isotropic chiral medium and the fractional curl operator in electromagnetics. Currently, he is a doctoral student at
 the Pennsylvania State University. He is a student member of SPIE.

\vspace{2ex}\noindent{\bf John A. Polo Jr.} received a B.S. degree in Physics from the University of
Massachusetts, Amherst in 1973 and a Ph.D. degree in Physics from the University of Virginia in 1979.  He has held his
present position at Edinboro University of Pennsylvania since 1990.  His research in the past was conducted in
experimental condensed matter physics.  Since 2000 he has been working in electromagnetics theory with current
research interests in the optical properties of complex materials and metamaterials.  He is a member of SPIE.

\vspace{2ex}\noindent{\bf Akhlesh Lakhtakia} received degrees from the Banaras Hindu University (B.Tech. \& D.Sc.)
and the University of Utah (M.S. \& Ph.D.), in Electronics Engineering and Electrical
Engineering, respectively. He is the Charles Godfrey Binder (Endowed) Professor of Engineering
Science and Mech\-anics at the Pennsylvania State University.  His current research interests include nanotechnology, plasmonics,
complex materials, metamaterials, and sculptured thin films. He is a Fellow of SPIE, OSA, AAAS, and the
Institute of Physics (UK).


\begin{thebibliography}{99}

\bibitem[1]{Borstel}
 G. Borstel and H. J. Falge, ``Surface phonon-polaritons," in \emph{Electromagnetic Surface Modes},
A. D. Boardman, Ed., Chap. 6, Wiley, New York, NY, USA (1982).

\bibitem[2]{Wallis}
R. F. Wallis, ``Surface magnetoplasmons on
semiconductors," in \emph{Electromagnetic Surface Modes},
A. D. Boardman, Ed., Chap. 15, Wiley, New York, NY, USA (1982).

\bibitem[3]{Elston}
S. J. Elston and J. R. Sambles,
``Surface plasmon-polaritons on an anisotropic substrate,"
\emph{J. Mod. Opt.} {\bf 37}, 1895-1902 (1990)
[doi:10.1080/09500349014552101].

\bibitem[4]{Depine95}
R. A. Depine and M. L. Gigli,
``Excitation of surface plasmons and total absorption of light at
the flat boundary between a metal and a uniaxial crystal,"
\emph{Opt. Lett.} {\bf 20}, 2243-2245 (1995)
[doi:10.1364/OL.20.002243].

\bibitem[5]{Wang}
H. Wang,
``Excitation of surface plasmon oscillations
at an interface between anisotropic dielectric and metallic media,"
\emph{Opt. Mater.} {\bf 4}, 651-656 (1995)
[doi:10.1016/0925-3467(95)00013-5].

\bibitem[6]{Depine97}
R. A. Depine and M. L. Gigli,
``Resonant excitation of surface modes at a single
flat uniaxial-metal interface,"
\emph{J. Opt. Soc. Am. A} {\bf 14}, 510-519 (1997)
[doi:10.1364/JOSAA.14.000510].

\bibitem[7]{Yan}
W. Yan, L. Shen, L. Ran, and J. A. Kong,
``Surface modes at the interfaces between isotropic
media and indefinite media,"
\emph{J. Opt. Soc. Am. A} {\bf 24}, 530-535 (2007)
[doi:10.1364/JOSAA.24.000530].

\bibitem[8]{Abdulhalim}
I. Abdulhalim,
``Surface plasmon TE and TM waves at the
anisotropic film-metal interface,"
\emph{J. Opt. A: Pure Appl. Opt.} {\bf 11}, 015002 (2009)
[doi:10.1088/1464-4258/11/1/015002].

\bibitem[9]{PL2009}
J. A. Polo Jr. and A.~Lakhtakia,
 ``On the surface plasmon polariton wave at the planar interface of a
 metal and a chiral sculptured thin film,"
 \emph{Proc. R. Soc. Lond. A\/}  {\bf 465}, 87-107
(2009) [doi:10.1098/rspa.2008.0211].

\bibitem[10]{PLjosaa}
J. A. Polo, Jr. and A.~Lakhtakia,
``Energy flux in a surface-plasmon-polariton wave
bound to the planar interface of a metal
and a structurally chiral material,"
 \emph{J. Opt. Soc. Am. A\/}  {\bf 26}, 1696-1703
(2009) [doi:10.1364/JOSAA.26.001696].

\bibitem[11]{Sprokel_I}
G. J. Sprokel, R. Santo, and J. D. Swalen,
``Determination of the surface tilt angle by attenuated
total reflection,"
\emph{Mol. Cryst. Liq. Cryst.} {\bf 68}, 29-38 (1981)
[doi:10.1080/00268948108073550].

\bibitem[12]{Sprokel_II}
G. J. Sprokel,
``The reflectivity of a liquid crystal cell in a surface plasmon experiment,"
\emph{Mol. Cryst. Liq. Cryst.} {\bf 68}, 39-45 (1981)
[doi:10.1080/00268948108073551].

\bibitem[13]{Gaspar}
J. A. Gaspar-Armenta and F. Villa,
``Photonic surface-wave excitation: photonic
crystal-metal interface,"
\emph{J. Opt. Soc. Am. B} {\bf 20}, 2349-2354 (2003)
[doi:10.1364/JOSAB.20.002349].

\bibitem[14]{Das}
R. Das and R. Jha,
``On the modal characteristics of surface plasmon polaritons
at a metal-Bragg interface at optical frequencies,"
\emph{Appl. Opt.} {\bf 48}, 4904-4908 (2009)
[doi:10.1364/AO.48.004904].

\bibitem[15]{Guo_I}
J. Guo and R. Adato,
``Extended long range plasmon waves in finite
thickness metal film and layered dielectric
materials,"
\emph{Opt. Express} {\bf 14}, 12409-12418 (2006)
[doi:10.1364/OE.14.012409].

\bibitem[16]{Guo_II}
R. Adato and J. Guo,
``Characteristics of ultra-long range surface
plasmon waves at optical frequencies,"
\emph{Opt. Express} {\bf 15}, 5008-5017 (2007)
[doi:10.1364/OE.15.005008].

\bibitem[17]{PartI}
M. A. Motyka and A.~Lakhtakia,
 ``Multiple trains of same-color surface plasmon-polaritons guided by the planar interface of a metal and a sculptured nematic thin film,"
 \emph{J. Nanophoton.\/}   {\bf 2}, 021910 (2008) [doi:10.1117/1.3033757].


 \bibitem[18]{PartII}
M. A. Motyka and A.~Lakhtakia,
 ``Multiple trains of same-color surface plasmon-polaritons guided
 by the planar interface of a metal and a sculptured nematic thin film. Part~II: Arbitrary incidence,"
 \emph{J. Nanophoton.\/} {\bf 3}, 033502 (2009) [doi:10.1117/1.3147876].

 \bibitem[19]{PartIII}
A.~Lakhtakia, Y.-J. Jen, and C.-F. Lin,
 ``Multiple trains of same-color surface plasmon-polaritons guided
 by the planar interface of a metal and a sculptured nematic thin film. Part~III: Experimental evidence,"
 \emph{J. Nanophoton.\/} {\bf 3}, 033506 (2009) [doi:10.1117/1.3249629].


\bibitem[20]{DPL2009}
Devender, D.~P.~Pulsifer and A.~Lakhtakia,
 ``Multiple surface plasmon polariton waves,"
 \emph{Electron. Lett.\/}  {\bf 45},  1137-1138
(2009) [doi:10.1049/el.2009.2049].


\bibitem[21]{Lmsec}
A. Lakhtakia,  ``Sculptured thin films: accomplishments and emerging uses," \emph{Mater. Sci. Engg. C\/} {\bf 19},
427-434 (2002)
[doi:10.1016/S0928-4931(01)00438-6].

\bibitem[22]{LMbook}
A. Lakhtakia  and  R. Messier, \emph{Sculptured Thin Films: Nanoengineered
Morphology and Optics\/}, SPIE Press, Bellingham, WA, USA (2005).


 \bibitem[23]{APL2009}
K. Agarwal, J.~A. Polo Jr., and A. Lakhtakia,
``Theory of Dyakonov--Tamm waves at the
planar interface of a sculptured nematic
thin film and an isotropic dielectric
material,"
\emph{J. Opt. A: Pure Appl. Opt.\/} {\bf 11}, 074003 (2009)
[doi:10.1088/1464-4258/11/7/074003].

\bibitem[24]{YS75}
V.~A.~Yakubovich and V.~M.~Starzhinskii,
{\em Linear Differential Equations with Periodic Coefficients\/},
Wiley, New York, NY, USA (1975).

\bibitem[25]{HWH}
I.~J. Hodgkinson, Q.~h. Wu, and J. Hazel,
``Empirical equations for the principal refractive
indices and column angle of obliquely deposited films of tantalum
oxide, titanium oxide, and zirconium oxide,"
\emph{Appl. Opt.\/} {\bf  37}, 2653-2659 (1998)
[doi:10.1364/AO.37.002653].


\bibitem[26]{Jaluria}
Y.~Jaluria,
{\em Computer Methods for Engineering\/},
{Taylor \& Francis}, Washington, DC, USA (1996).


 \end{thebibliography}
 \end{document}